\documentclass[11pt]{article}
\input{psfig.sty}
\newcommand{\dshv}{
\begin{picture}(52,10)
\put(15,3){\line(-5,2){10}}
\put(15,3){\line(-5,-2){10}}
\multiput(15,3)(22,0){2}{\circle*{2}}
\multiput(15,3)(6,0){4}{\line(1,0){4}}
\put(37,3){\line(5,2){10}}
\put(37,3){\line(5,-2){10}}
\end{picture}  }

\newcommand{\zigv}{
\begin{picture}(50,10)
\put(15,3){\line(-5,2){10}}
\put(15,3){\line(-5,-2){10}}
\multiput(15,3)(20.15,0){2}{\circle*{2}}
\multiput(15,3)(0.42,0.42){4}{\circle*{1}}
\multiput(16.26,4.25)(0.5,-0.5){6}{\circle*{1}}
\multiput(18.76,1.75)(0.5,0.5){6}{\circle*{1}}
\multiput(21.26,4.25)(0.5,-0.5){6}{\circle*{1}}
\multiput(23.76,1.75)(0.5,0.5){6}{\circle*{1}}
\multiput(26.26,4.25)(0.5,-0.5){6}{\circle*{1}}
\multiput(28.76,1.75)(0.5,0.5){6}{\circle*{1}}
\multiput(31.26,4.25)(0.5,-0.5){6}{\circle*{1}}
\multiput(33.76,1.75)(0.42,0.42){4}{\circle*{1}}
\put(35.15,3){\line(5,2){10}}
\put(35.15,3){\line(5,-2){10}}
\end{picture}  }
\newcommand{\dotv}{
\begin{picture}(33,10)
\multiput(6,3)(4,0){3}{\circle*{2}}
\put(18,3){\circle*{3}}
\put(18,3){\line(5,2){10}}
\put(18,3){\line(5,-2){10}}
\end{picture}  }
\newcommand{\ddotv}{
\begin{picture}(54,10)
\put(15,3){\line(-5,2){10}}
\put(15,3){\line(-5,-2){10}}
\multiput(15,3)(24,0){2}{\circle*{3}}
\multiput(19,3)(4,0){5}{\circle*{2}}
\put(39,3){\line(5,2){10}}
\put(39,3){\line(5,-2){10}}
\end{picture}  }
\newcommand{\Gausd}{
\begin{picture}(55,10)
\thicklines
\put(6,-1){\line(1,0){40}}
\put(6,1){{\bf k}}
\put(35,1){-{\bf k}}
\end{picture}  }
\newcommand{\dashd}{
\begin{picture}(66,15)
\thicklines
\put(6,-1){\line(1,0){54}}
\put(6,1){{\bf k}}
\put(49,1){-{\bf k}}
\multiput(18,-1)(28,0){2}{\circle*{3}}
\multiput(18,-1)(0.05,0.5){9}{\circle*{1}}
\multiput(46,-1)(-0.05,0.5){9}{\circle*{1}}
\multiput(19,5)(0.3,0.475){9}{\circle*{1}}
\multiput(45,5)(-0.3,0.475){9}{\circle*{1}}
\multiput(23.4,10.6)(0.45,0.25){9}{\circle*{1}}
\multiput(40.6,10.6)(-0.45,0.25){9}{\circle*{1}}
\multiput(30,13.2)(0.25,0){17}{\circle*{1}}
\end{picture}  }
\newcommand{\zigd}{
\begin{picture}(66,15)
\thicklines
\put(6,-1){\line(1,0){54}}
\put(6,1){{\bf k}}
\put(49,1){-{\bf k}}
\multiput(18,-1)(26.6,0){2}{\circle*{3}}
\multiput(18,-1)(-0.1,0.4){11}{\circle*{1}}
\multiput(44.6,-1)(0.1,0.4){11}{\circle*{1}}
\multiput(17,3)(0.4,-0.05){11}{\circle*{1}}
\multiput(45.6,3)(-0.4,-0.05){11}{\circle*{1}}
\multiput(21,2.5)(-0.1,0.4){11}{\circle*{1}}
\multiput(41.6,2.5)(0.1,0.4){11}{\circle*{1}}
\multiput(20,6.5)(0.4,-0.1){11}{\circle*{1}}
\multiput(42.6,6.5)(-0.4,-0.1){11}{\circle*{1}}
\multiput(24,5.5)(0,0.4){11}{\circle*{1}}
\multiput(38.6,5.5)(0,0.4){11}{\circle*{1}}
\multiput(24,9.5)(0.35,-0.22){11}{\circle*{1}}
\multiput(38.6,9.5)(-0.35,-0.22){11}{\circle*{1}}
\multiput(27.5,7.3)(0.15,0.37){11}{\circle*{1}}
\multiput(35.1,7.3)(-0.15,0.37){11}{\circle*{1}}
\multiput(29,11)(0.23,-0.32){11}{\circle*{1}}
\multiput(33.6,11)(-0.23,-0.32){11}{\circle*{1}}
\end{picture}  }
\newcommand{\dotd}{
\begin{picture}(66,15)
\thicklines
\put(6,-1){\line(1,0){54}}
\put(6,1){{\bf k}}
\put(49,1){-{\bf k}}
\multiput(18,-1)(28,0){2}{\circle*{3}}
\multiput(18.5,3)(27,0){2}{\circle*{2}}
\multiput(20.6,7)(22.8,0){2}{\circle*{2}}
\multiput(24,10)(16,0){2}{\circle*{2}}
\multiput(28,12)(8,0){2}{\circle*{2}}
\put(32,12.6){\circle*{2}}
\end{picture}  }
\newcommand{\bldot}{
\begin{picture}(32,12)
\put(9,3.5){\circle{11}}
\multiput(8.8,-1.6)(0.5,0.5){12}{\circle*{1}}
\multiput(9.2,8.6)(-0.5,-0.5){12}{\circle*{1}}
\multiput(6.2,-1)(0.5,0.5){15}{\circle*{1}}
\multiput(11.8,8)(-0.5,-0.5){15}{\circle*{1}}
\multiput(14.5,3.5)(3.5,0){5}{\circle*{2}}
\end{picture}  }
\newcommand{\blddot}{
\begin{picture}(28.5,12)
\put(9,3.5){\circle{11}}
\multiput(8.8,-1.6)(0.5,0.5){12}{\circle*{1}}
\multiput(9.2,8.6)(-0.5,-0.5){12}{\circle*{1}}
\multiput(6.2,-1)(0.5,0.5){15}{\circle*{1}}
\multiput(11.8,8)(-0.5,-0.5){15}{\circle*{1}}
\multiput(14.3,6)(3.5,0.75){4}{\circle*{2}}
\multiput(14.3,1)(3.5,-0.75){4}{\circle*{2}}
\end{picture}  }
\newcommand{\bltdot}{
\begin{picture}(28.5,15)
\put(9,3.5){\circle{11}}
\multiput(8.8,-1.6)(0.5,0.5){12}{\circle*{1}}
\multiput(9.2,8.6)(-0.5,-0.5){12}{\circle*{1}}
\multiput(6.2,-1)(0.5,0.5){15}{\circle*{1}}
\multiput(11.8,8)(-0.5,-0.5){15}{\circle*{1}}
\multiput(14,7)(3.5,0.8){4}{\circle*{2}}
\multiput(14,0)(3.5,-0.8){4}{\circle*{2}}
\multiput(14.5,3.5)(3.5,0){4}{\circle*{2}}
\end{picture}  }
\newcommand{\blzig}{
\begin{picture}(32.5,12)
\put(9,3.5){\circle{11}}
\multiput(8.8,-1.6)(0.5,0.5){12}{\circle*{1}}
\multiput(9.2,8.6)(-0.5,-0.5){12}{\circle*{1}}
\multiput(6.2,-1)(0.5,0.5){15}{\circle*{1}}
\multiput(11.8,8)(-0.5,-0.5){15}{\circle*{1}}
\multiput(14.5,3.5)(0.5,0.5){3}{\circle*{1}}
\multiput(15.5,4.5)(0.5,-0.5){5}{\circle*{1}}
\multiput(17.5,2.5)(0.5,0.5){5}{\circle*{1}}
\multiput(19.5,4.5)(0.5,-0.5){5}{\circle*{1}}
\multiput(21.5,2.5)(0.5,0.5){5}{\circle*{1}}
\multiput(23.5,4.5)(0.5,-0.5){5}{\circle*{1}}
\multiput(25.5,2.5)(0.5,0.5){5}{\circle*{1}}
\multiput(27.5,4.5)(0.5,-0.5){5}{\circle*{1}}
\end{picture}  }
\newcommand{\bldzig}{
\begin{picture}(30,12)
\put(9,3.5){\circle{11}}
\multiput(8.8,-1.6)(0.5,0.5){12}{\circle*{1}}
\multiput(9.2,8.6)(-0.5,-0.5){12}{\circle*{1}}
\multiput(6.2,-1)(0.5,0.5){15}{\circle*{1}}
\multiput(11.8,8)(-0.5,-0.5){15}{\circle*{1}}
\multiput(14.2,5.5)(0.4,0.55){5}{\circle*{1}}
\multiput(15.8,7.7)(0.55,-0.4){5}{\circle*{1}}
\multiput(18,6.1)(0.4,0.55){5}{\circle*{1}}
\multiput(19.6,8.3)(0.55,-0.4){5}{\circle*{1}}
\multiput(21.8,6.7)(0.4,0.55){5}{\circle*{1}}
\multiput(23.4,8.9)(0.55,-0.4){5}{\circle*{1}}
\multiput(25.6,7.3)(0.32,0.44){4}{\circle*{1}}
\multiput(14.2,1.5)(0.4,-0.55){5}{\circle*{1}}
\multiput(15.8,-0.7)(0.55,0.4){5}{\circle*{1}}
\multiput(18,0.9)(0.4,-0.55){5}{\circle*{1}}
\multiput(19.6,-1.3)(0.55,0.4){5}{\circle*{1}}
\multiput(21.8,0.3)(0.4,-0.55){5}{\circle*{1}}
\multiput(23.4,-1.9)(0.55,0.4){5}{\circle*{1}}
\multiput(25.6,-0.3)(0.32,-0.44){4}{\circle*{1}}
\end{picture}  }
\newcommand{\bltzig}{
\begin{picture}(30,15)
\put(9,3.5){\circle{11}}
\multiput(8.8,-1.6)(0.5,0.5){12}{\circle*{1}}
\multiput(9.2,8.6)(-0.5,-0.5){12}{\circle*{1}}
\multiput(6.2,-1)(0.5,0.5){15}{\circle*{1}}
\multiput(11.8,8)(-0.5,-0.5){15}{\circle*{1}}
\multiput(13.7,7)(0.4,0.55){5}{\circle*{1}}
\multiput(15.3,9.2)(0.55,-0.4){5}{\circle*{1}}
\multiput(17.5,7.6)(0.4,0.55){5}{\circle*{1}}
\multiput(19.1,9.8)(0.55,-0.4){5}{\circle*{1}}
\multiput(21.3,8.2)(0.4,0.55){5}{\circle*{1}}
\multiput(22.9,10.4)(0.55,-0.4){5}{\circle*{1}}
\multiput(25.1,8.8)(0.32,0.44){4}{\circle*{1}}
\multiput(13.7,0)(0.4,-0.55){5}{\circle*{1}}
\multiput(15.3,-2.2)(0.55,0.4){5}{\circle*{1}}
\multiput(17.5,-0.6)(0.4,-0.55){5}{\circle*{1}}
\multiput(19.1,-2.8)(0.55,0.4){5}{\circle*{1}}
\multiput(21.3,-1.2)(0.4,-0.55){5}{\circle*{1}}
\multiput(22.9,-3.4)(0.55,0.4){5}{\circle*{1}}
\multiput(25.1,-1.8)(0.32,-0.44){4}{\circle*{1}}
\multiput(14.5,3.5)(0.5,0.5){3}{\circle*{1}}
\multiput(15.5,4.5)(0.5,-0.5){5}{\circle*{1}}
\multiput(17.5,2.5)(0.5,0.5){5}{\circle*{1}}
\multiput(19.5,4.5)(0.5,-0.5){5}{\circle*{1}}
\multiput(21.5,2.5)(0.5,0.5){5}{\circle*{1}}
\multiput(23.5,4.5)(0.5,-0.5){5}{\circle*{1}}
\multiput(25.5,2.5)(0.5,0.5){3}{\circle*{1}}
\end{picture}  }
\newcommand{\selfe}{
\begin{picture}(48,10)
\thicklines
\put(24,3.5){\circle{11}}
\multiput(23.8,-1.6)(0.5,0.5){12}{\circle*{1}}
\multiput(24.2,8.6)(-0.5,-0.5){12}{\circle*{1}}
\multiput(21.2,-1)(0.5,0.5){15}{\circle*{1}}
\multiput(26.8,8)(-0.5,-0.5){15}{\circle*{1}}
\put(18.5,3.5){\line(-1,0){15}}
\put(29.5,3.5){\line(1,0){15}}
\end{picture}  }

\begin{document}

\title{
\textbf{Perturbation theory methods applied to critical phenomena}
}

\author{J. Kaupu\v{z}s
\thanks{E--mail: \texttt{kaupuzs@latnet.lv}} \\
Institute of Mathematics and Computer Science, University of Latvia\\
29 Rainja Boulevard, LV--1459 Riga, Latvia}

\date{\today}

\maketitle

\begin{abstract}
Different perturbation theory treatments of the Ginzburg--Landau phase
transition model are discussed. This includes a criticism of the
perturbative renormalization group (RG) approach and a proposal of
a novel method providing critical exponents consistent with the
known exact solutions in two dimensions. The new
values of critical exponents are discussed and compared to the results
of numerical simulations and experiments.
\end{abstract}


\section{Introduction}
Phase transitions and critical phenomena is one of the most widely
investigated topics in modern physics. Nevertheless, a limited
number of exact and rigorous results is available~\cite{Baxter}.
Our purpose is to give a critical analysis of the conventional approach
in calculation of critical exponents based on the perturbative
renormalization group (RG) theory~\cite{Wilson,Ma,Justin} and to propose
a new method which provides results consistent with the known exact solutions.
The basic hypothesis of the conventional (RG) theory is the existence of
a certain fixed point for the RG transformation. However, the existence of
such a stable fixed point for the Ginzburg--Landau model (which
lies in the basis of the field theory) has not been proven
mathematically in the case of the spatial dimensionality $d<4$.

 The usual RG theory treatment of the Ginzburg--Landau model
is based on the diagrammatic perturbation theory (Feynman diagrams).
We have demonstrated that this treatment is
contradictory and therefore cannot give correct values of critical
exponents. Namely, based on a method which is mathematically correct
and well justified in view of the conventional RG theory, we prove
the nonexistence of the non--Gaussian fixed point predicted by this theory
(Sect.~\ref{sec:RG}). In Sect.~\ref{sec:random} we prove that a correctly
treated diagram expansion provides results which essentially
differ from those of the perturbative (diagrammatic) RG theory.
Finally, we have proposed a novel analytical method of
determination of critical exponents in the Ginzburg--Landau model
(Sect.~\ref{sec:my}), and have compared the predicted exact values of
critical exponents to the results of numerical and real experiments
(Sect.~\ref{sec:compare}).

\section{Critical analysis of the perturbative RG method} \label{sec:RG}

Here we consider the Ginzburg--Landau phase transition model within
the usual renormalization group approach to show that this approach
is contradictory (for more details see also~\cite{kau_cm1}).
The Hamiltonian of this model in the Fourier representation reads
\begin{equation} \label{eq:H}
\frac{H}{T}= \sum\limits_{\bf k} \left( r_0+c \,{\bf k}^2 \right)
{\mid \varphi_{\bf k} \mid}^2 + uV^{-1}
\sum\limits_{{\bf k}_1,{\bf k}_2,{\bf k}_3}
\varphi_{{\bf k}_1} \varphi_{{\bf k}_2} \varphi_{{\bf k}_3}
\varphi_{-{\bf k}_1-{\bf k}_2-{\bf k}_3} \;,
\end{equation}
where $\varphi_{\bf k}=V^{-1/2} \int \varphi({\bf x})
\exp(-i {\bf kx}) \, d{\bf x}$ are Fourier components of the scalar
order parameter field $\varphi({\bf x})$, $T$ is the temperature,and
$V$ is the volume of the system. In the RG field theory~\cite{Ma,Justin}
Hamiltonian (\ref{eq:H}) is renormalized by integration of
$\exp(-H/T)$ over $\varphi_{\bf k}$ with $\Lambda/s<k<\Lambda$,
followed by a certain rescaling procedure providing a Hamiltonian
corresponding to the initial values of $V$ and $\Lambda$, where
$\Lambda$ is the upper cutoff of the
$\varphi^4$ interaction. Due to this procedure, additional terms
appear in the Hamiltonian (\ref{eq:H}), so that in general the
renormalized Hamiltonian contains a continuum of parameters.
The basic hypothesis of the RG theory in $d<4$ dimensions is the
existence of a non--Gaussian fixed point $\mu=\mu^*$ for the RG
transformation $R_s$ defined in the space of Hamiltonian parameters, i.e.,
\begin{equation} \label{eq:fixp}
R_s \mu^* = \mu^* \;.
\end{equation}
The fixed-point values of the Hamiltonian parameters are marked by an
asterisk ($r_0^*$, $c^*$, and $u^*$, in particular). Note that
$\mu^*$ is unambiguously defined by fixing the values of $c^*$
and $\Lambda$. According to the RG
theory, the main terms in the renormalized Hamiltonian in
$d=4-\epsilon$ dimensions are those contained in (\ref{eq:H}) with
$r_0^*$ and $u^*$ of the order $\epsilon$,
whereas the additional terms are small corrections of order $\epsilon^2$.

Consider the Fourier transform $G({\bf k}, \mu)$ of the two--point
correlation (Green's) function, corresponding to a point $\mu$.
Under the RG transformation $R_s$ this function transforms as
follows~\cite{Ma}
\begin{equation} \label{eq:RGt}
G({\bf k}, \mu)=s^{2- \eta} \, G(s {\bf k}, R_s \mu) \;.
\end{equation}
Let $G({\bf k}, \mu) \equiv G(k,\mu)$ \,
(at ${\bf k} \ne {\bf0}$ and $V \to \infty$)
be defined within $k \le \Lambda$.
Since Eq.~(\ref{eq:RGt}) holds for any
$s>1$, we can set $s= \Lambda/k$, which at $\mu = \mu^*$ yields
\begin{equation} \label{eq:asyfix}
G({\bf k},\mu^*) = a \, k^{-2 + \eta} \hspace{3ex}
\mbox{for} \;\; k<\Lambda \;,
\end{equation}
where $a= \Lambda^{2- \eta} G(\Lambda, \mu^*)$ is the amplitude
and $\eta$ is the universal critical exponent. According to the
universality hypothesis, the infrared behavior of the Green's
function is described by the same universal value of $\eta$ at
any $\mu$ on the critical surface (with the only requirement that
all parameters of Hamiltonian (\ref{eq:H}) are present), i.e.,
\begin{equation} \label{eq:asy}
G({\bf k},\mu)= b(\mu) \, k^{-2+\eta} \hspace*{3ex}
\mbox{at} \;\; k \to 0\;,
\end{equation}
where
\begin{equation} \label{eq:lim}
b(\mu)=\lim\limits_{k \to 0} k^{2-\eta} \, G({\bf k},\mu) \;.
\end{equation}
According to Eq.~(\ref{eq:RGt}), which holds for any $s=s(k)>1$
and for $s=\Lambda/k$ in particular,
Eq.~(\ref{eq:lim}) reduces to
\begin{equation} \label{eq:b}
b(\mu)=\lim\limits_{k \to 0} k^{2-\eta} s(k)^{2-\eta} \,
G(s {\bf k},R_s \mu) =a \;,
\end{equation}
if the fixed point
$\mu^* = \lim\limits_{s \to \infty} R_s \mu$ exists.
Let us define the function $X({\bf k},\mu)$
as $X({\bf k},\mu) \, = \, k^{-2} G^{-1} ({\bf k},\mu)$.
According to Eqs.~(\ref{eq:asyfix}), (\ref{eq:asy}), and
(\ref{eq:b}), we have (for $k< \Lambda$)
\begin{equation} \label{eq:Xfix}
X({\bf k},\mu^*)= \frac{1}{a} \, k^{-\eta}
\end{equation}
and
\begin{equation} \label{eq:Xcrit}
X({\bf k},\mu)= \frac{1}{a} \, k^{-\eta} \,
+ \, \delta X({\bf k},\mu) \;,
\end{equation}
where $\mu$ belongs to the critical surface,
and $\delta X({\bf k},\mu)$ denotes the correction--to--scaling term.
From (\ref{eq:Xfix}) and (\ref{eq:Xcrit}) we obtain the equation
\begin{equation} \label{eq:eq}
\delta X({\bf k},\mu^* + \delta \mu) =
X({\bf k}, \mu^* + \delta \mu) - X({\bf k}, \mu^*) \; ,
\end{equation}
where $\delta \mu= \mu - \mu^*$.
Since Eq.~(\ref{eq:eq}) is true for any small deviation $\delta \mu$
satisfying the relation
\begin{equation} \label{eq:ff}
\mu^* =\lim\limits_{s \to \infty} R_s(\mu^* +\delta\mu) \;,
\end{equation}
we choose $\delta \mu$ such that $\mu^* \Rightarrow \mu^* +
\delta\mu$ corresponds to the variation of the Hamiltonian parameters
 $r_0^* \Rightarrow r_0^* + \delta r_0$,
$c^* \Rightarrow c^* + \delta c$, and
$u^* \Rightarrow u^* + \,\epsilon \times \Delta$,
where $\Delta$ is a small constant.
The values of $\delta r_0$ and $\delta c$ are choosen to fit the
critical surface and to meet the condition (\ref{eq:ff}) at fixed
$c^*=1$ and $\Lambda=1$. In particular, quantity $\delta c$ is found
$\delta c \,=\, B \; \epsilon^2 \, +o(\epsilon^3)$
with some (small) coefficient $B=B(\Delta)$, to compensate the shift in
$c$ of the order $\epsilon^2$ due to the renormalization (cf.~\cite{Ma}).
The formal $\epsilon$--expansion of $\delta X({\bf k}, \mu)$,
defined by Eq.~(\ref{eq:eq}), can be obtained in the usual way from the
perturbation theory. This yields
\begin{equation} \label{eq:expas}
\delta X({\bf k}, \mu) = \epsilon^2 \, [\, C_1(\Delta) +
C_2(\Delta) \, \ln k \, ] \, +o(\epsilon^3) \hspace*{3ex}
\mbox{at} \;\; k \to 0 \;,
\end{equation}
where $C_1(\Delta)$ and $C_2(\Delta)$ ($C_2 \ne 0$) are coefficients
independent on $\epsilon$.

It is commonly accepted in the RG field theory to make an
expansion like (\ref{eq:expas}), obtained from the diagrammatic
perturbation theory, to fit an asymptotic expansion in $k$ powers,
thus determining the critical exponents.
In general, such a method is not rigorous since,
obviously, there exist such functions which do not contribute
to the asymptotic expansion in $k$ powers at $k \to 0$, but give a
contribution to the formal $\epsilon$--expansion at any fixed
$k$. Besides, the expansion coefficients do not vanish at $k \to 0$.
Trivial examples of such functions are
$\epsilon^m \, \exp(-\epsilon k^{-\epsilon})$
and $\epsilon^m \,[1-\tanh(\epsilon \,k^{-\epsilon})]$ where $m$ is integer.
Nevertheless, according to the general ideas of the RG theory
(not based on Eq.~(\ref{eq:eq})),
in the vicinity of the fixed point the asymptotic expansion
\begin{equation} \label{eq:Xas}
X({\bf k},\mu)= \frac{1}{a} k^{-\eta} +
b_1 k^{\epsilon+o(\epsilon^2)} + b_2 k^{2+o(\epsilon)} + ...
\end{equation}
is valid not only at $k \to 0$, but within $k<\Lambda$. The latter means
that terms of the kind $\epsilon^m \, \exp(-\epsilon k^{-\epsilon})$
are absent or negligible.
Thus, if the fixed point does exist, then we can
obtain correct $\epsilon$--expansion of $\delta X({\bf k},\mu)$
at small $k$ by expanding the term $b_1 k^{\epsilon+o(\epsilon^2)}$
(with $b_1=b_1(\epsilon,\Delta)$) in
Eq.~(\ref{eq:Xas}) in $\epsilon$ powers, and the result must
agree with (\ref{eq:expas}) at small $\Delta$, at least.
The latter, however, is impossible
since Eq.~(\ref{eq:expas}) never agree with
\begin{equation}
\delta X({\bf k},\mu) = b_1(\epsilon,\Delta) \,
\left[\, 1+ \epsilon \ln k +o(\epsilon^2) \, \right]
\end{equation}
obtained from (\ref{eq:Xas}) at $k \to 0$. Thus,
in its very basics the perturbative RG method in $4-\epsilon$ dimensions
is contradictory. From this we can conclude that the initial
assumption about existence of a certain fixed point, predicted
by the RG field theory in $4-\epsilon$ dimensions, is not valid.

\section{A model with quenched randomness} \label{sec:random}

Here we consider the Ginzburg--Landau phase transition model with $O(n)$
symmetry (i.e., the $n$--vector model) which includes a quenched randomness,
i.e., a random temperature disorder (for more details see
also~\cite{kau_cm2}). One of the basic ideas of the
perturbative RG theory is that $n$ may be considered as a continuous
parameter and the limit $n \to 0$  makes sense describing the self--avoiding
random walk or statistics of polymers~\cite{Ma,Justin}.
We have proven rigorously that within the diagrammatic perturbation theory
the quenched randomness does not change the critical exponents at $n \to 0$,
which is in contrast to the prediction of the conventional RG theory
formulated by means of the Feynman diagrams.

 The Hamiltonian of the actually considered model is
\begin{eqnarray} \label{eq:Hr}
H/T &=& \int \left[ \left(r_0+ \sqrt{u} \, f({\bf x}) \right)
\varphi^2({\bf x}) + c \, (\nabla \varphi({\bf x}) )^2 \right] d{\bf x}
\\
&+& \, uV^{-1} \sum\limits_{i,j,{\bf k}_1,{\bf k}_2,{\bf k}_3 }
\varphi_i({\bf k}_1) \varphi_i({\bf k}_2) \, u_{{\bf k}_1+{\bf k}_2} \,
\varphi_j({\bf k}_3) \varphi_j(-{\bf k}_1-{\bf k}_2-{\bf k}_3) \nonumber
\end{eqnarray}
which includes a random temperature (or random mass) disorder
represented by the term $\sqrt{u}\,f({\bf x})\,\varphi^2({\bf x})$.
For convenience, we call this model the random model. In Eq.~(\ref{eq:Hr})
$\varphi({\bf x})$ is an $n$--component vector with components
$\varphi_i({\bf x})=V^{-1/2} \sum_{k<\Lambda} \varphi_i({\bf k})
e^{i{\bf kx}}$,
depending on the coordinate ${\bf x}$, and
$f({\bf x})=V^{-1/2} \sum_{\bf k} f_{\bf k} e^{i{\bf kx}}$ is a random
variable with the Fourier components
$f_{\bf k}=V^{-1/2} \int f({\bf x}) e^{-i{\bf kx}} d{\bf x}$.

The system is characterized by the two--point correlation function
$G_i({\bf k})$ defined by the equation
\begin{equation}
\left< \varphi_i({\bf k}) \varphi_j(-{\bf k}) \right>
= \delta_{i,j} \, G_i({\bf k}) = \delta_{i,j} \, G({\bf k})\;.
\end{equation}
It is supposed that the averaging is performed over the $\varphi({\bf x})$
configurations and then over the $f({\bf x})$ configurations with a fixed
(quenched) Gaussian distribution $P(\{ f_{\bf k} \})$ for the set of Fourier
components $\{ f_{\bf k} \}$,
i.~e., our random model describes a quenched randomness.

We have proven the following theorem.
\vspace*{1ex}

{\it Theorem}. \, In the limit $n \to 0$, the perturbation expansion
of the correlation function  $G({\bf k})$ in  $u$  power series for the
random model with the Hamiltonian (\ref{eq:Hr}) is identical to the
perturbation expansion for the corresponding model with the Hamiltonian
\begin{eqnarray} \label{eq:Hp}
H/T &=& \int \left[ r_0 \, \varphi^2({\bf x})
+ c \, (\nabla \varphi({\bf x}) )^2 \right] d{\bf x} \\
&+& \, uV^{-1} \sum\limits_{i,j,{\bf k}_1,{\bf k}_2,{\bf k}_3 }
\varphi_i({\bf k}_1) \varphi_i({\bf k}_2) \, \tilde u_{{\bf k}_1+{\bf k}_2} \,
\varphi_j({\bf k}_3) \varphi_j(-{\bf k}_1-{\bf k}_2-{\bf k}_3) \nonumber
\end{eqnarray}
where $\tilde u_{\bf k}=u_{\bf k}- {1 \over 2}
\left< {\mid f_{\bf k} \mid}^2 \right>$. \medskip

   For convenience, we call the model without the term
$\sqrt{u} \, f({\bf x}) \, \varphi^2({\bf x})$ the pure model, since this
term simulates the effect of random impurities~\cite{Ma}.
\vspace*{1ex}

 {\it Proof of the theorem}. \, According to the rules of the diagram
technique, the formal expansion for $G({\bf k})$ involves all connected
diagrams with two fixed outer solid lines. In the case of the pure model,
diagrams are constructed of the vertices {\mbox \zigv,} with factor
$-uV^{-1} \tilde u_{\bf k}$ related to any zigzag line with wave vector
${\bf k}$.
The solid lines are related to the correlation function in the Gaussian
approximation $G_0({\bf k})=1/ \left(2r_0+2ck^2 \right)$. Summation over
the components $\varphi_i({\bf k})$ of the vector $\varphi({\bf k})$
yields factor $n$ corresponding to each closed loop of solid lines in the
diagrams. According to this, the formal perturbation expansion is defined
at arbitrary $n$. In the limit $n \to 0$, all diagrams of $G({\bf k})$
vanish except those which do not contain the closed loops. In such a way,
for the pure model we obtain the expansion
\begin{equation} \label{eq:expu}
G({\bf k})= \Gausd + \, \zigd + \, ... \;.
\end{equation}
 In the case of the random model, the diagrams are constructed of
the vertices \dshv and {\mbox \dotv.} The factors
$uV^{-1} \left< {\mid f_{\bf k} \mid}^2 \right>$ correspond to the
coupled dotted lines and the factors $-uV^{-1} u_{\bf k}$ correspond to
the dashed lines. Thus, we have
\begin{equation} \label{eq:exra}
G({\bf k}) = \Gausd + \left[ \dashd + \dotd \right] + \, ... \;.
\end{equation}
In the random model, first the correlation function $G({\bf k})$
is calculated at a fixed $\{ f_{\bf k} \}$
(which corresponds to connected diagrams where solid lines are coupled, but the
dotted lines with factors $-\sqrt{u} \,V^{-1/2} f_{\bf k}$ are not coupled),
performing the averaging with the weight $P( \{ f_{\bf k} \})$ over the
configurations of the random variable
(i.e., the coupling of the dotted lines) afterwards.
According to this procedure, the diagrams of the random model in
general (not only at $n \to 0$) do not contain parts like {\mbox \bldot,}
{\mbox \blddot,} {\mbox \bltdot,} etc.,
which would appear only if unconnected (i.e., consisting of separate
parts) diagrams would be considered before the coupling of dotted lines.

It is evident from
Eqs.~(\ref{eq:expu}) and (\ref{eq:exra}) that all diagrams of the random
model are obtained from those of the pure model if any of the zigzag
lines is replaced either by a dashed or by a dotted line, performing
summation over all such possibilities. Such a method is valid in
the limit $n \to 0$, but not in general. The problem is that, except
the case $n \to 0$, the diagrams of the pure model contain parts like
{\mbox \blzig,} {\mbox \bldzig,} {\mbox \bltzig,} etc. If all the depicted
here zigzag lines are replaced by the dotted lines, then we obtain diagrams
which are not allowed in the random model, as explained
before. At $n \to 0$, the only problem is to determine the
combinatorial factors for the diagrams obtained by the above replacements.
For a diagram constructed of $M_1$ vertices \dshv and $M_2$ vertices \dotv
the combinatorial factor is the number of possible different couplings of
lines, corresponding to the given topological picture, divided by
$M_1 ! M_2 !$.

  Our further consideration is
valid also for the diagrams of free energy (at $n \to 0$
represented by the main terms containing single loop of solid lines)
and of $2m$--point correlation function.
We define that all diagrams which can be obtained from the
$i$--th diagram (i.e., the diagram of the $i$--th topology) of the pure
model, belong to the  $i$--th group. Obviously, all diagrams of the
$i$--th group represent a contribution of order $u^l$, where $l$ is
the total number of vertices \zigv in the $i$--th diagram.
The sum of the diagrams of the
$i$--th group can be found by the following algorithm.
\begin{itemize}
\item[1.]
Depict the $i$--th diagram of pure model in an a priori defined way.
\item[2.] Choose any one replacement
of the vertices \zigv by \dshv and {\mbox \ddotv,} and perform the summation
over all such possibilities. For any specific choice we consider only one
of the equivalent $M_1!M_2!$ distributions of the numbered $M_1$ vertices
\dshv and $M_2$ vertices \dotv over the fixed numbered positions instead of
the summation over all these distributions
with the weight $1/(M_1!M_2!)$. Thus, at this step the combinatorial factor
for any specific diagram is determined as
the number of possible distributions of lines (numbered before coupling)
for one fixed location of vertices consistent with the picture defined
in step 1.
\item[3.] The result of summation in
step 2  is divided by the number of independent symmetry transformations
(including the identical transformation) for the considered $i$--th diagram
constructed of vertices \zigv, since the same (original and transformed)
diagrams were counted as different.
\end{itemize}
   Note that the location of any vertex \dshv is defined by fixing the
position of dashed line, the orientation of which is not fixed. According
to this, the summation over all possible distributions of lines (numbered
before coupling) for one fixed location of vertices
yields factor $8^{M_1}4^{M_2/2}$. The $i$--th
diagram of the pure model also can be calculated by such an algorithm.
In this case we have $8^l$ line distributions, where $l=M_1+M_2/2$\,
is the total number of vertices \zigv in the  $i$--th diagram. Obviously,
the summation of diagrams of the  $i$--th  group can be performed with
factors $8^l$ instead of $8^{M_1}4^{M_2/2}$, but in this case
twice smaller factors
must be related to the coupled dotted lines.
The summation over all possibilities where zigzag lines are replaced by dashed
lines with factors $-uV^{-1}u_{\bf k}$ and by dotted lines with factors
${1 \over 2} uV^{-1} \left< {\mid f_{\bf k} \mid}^2 \right>$, obviously,
yields a factor
$uV^{-1} \left(-u_{\bf k}+{1 \over 2} \left< {\mid f_{\bf k} \mid}^2
\right> \right) \equiv -uV^{-1} \tilde u_{\bf k}$ corresponding to each
zigzag line with wave vector ${\bf k}$. Thus, the sum over the diagrams of
the  $i$--th group is identical to the  $i$--th diagram of the pure model
defined by Eq.~(\ref{eq:Hp}). By this the theorem has proved
not only for the two--point correlation function, but also for
$2m$--point correlation function and free energy.

If, in general, the factor $\sqrt{u}$ in Eq.~(\ref{eq:Hr}) is replaced by
$\sqrt{u'}$, where $u'$ is an independent expansion parameter,
then our analysis leads to the above relation between
diagrams for $u \, \tilde u_{\bf k}= u \, u_{\bf k}- {u'\over 2}
\left< {\mid f_{\bf k} \mid}^2 \right>$.
According to this, at $n \to 0$ the pure
and random models cannot be distinguished within the
diagrammatic perturbation theory. If, in principle, critical
exponents can be determined from the diagram expansions at
$n \to 0$, as it is suggested in the usual RG theory, then the same
critical exponents should be provided for both models at $n \to 0$.
In such a way, we conclude that the RG method is
not correct because the above condition is violated.
As compared to our simple treatment of the random model, the RG
treatment includes additional Feynman diagrams because the
Hamiltonian becomes more complicated after the renormalization.
However, this does not enable to find the
difference between both models: the original
information, when one starts the perturbative renormalization of
Hamiltonian~(\ref{eq:Hr}), is contained in the Feynman diagrams
we considered, but the renormalization by itself does not create
new information about the model. Really, by renormalization we merely
``forget'' some information about the short--wave fluctuations
to make that for the long--wave fluctuations easier accessible.
Thus, our conclusion remains true.

\section{New method based on perturbation theory} \label{sec:my}

 As we have already discussed in Sect.~\ref{sec:RG}, it is not a
rigorous method to make a formal expansion like (\ref{eq:expas})
and to try calculate the critical exponents therefrom. We
propose another treatment of the diagrammatic perturbation theory.
The basic idea is to obtain suitable equations by appropriate grouping
of the diagrams. Suitable are such equations which allow to find
the asymptotic expansions at the critical point directly in $k$
power series, but not in terms of the formal parameter $\ln k$
(as in Eq.~\ref{eq:expas}) which diverges at $k \to 0$. In such a
way, for the Ginzburg--Landau model defined by Eq.~(\ref{eq:Hp}),
where $u \, \tilde u_{\bf k}= u_{\bf k}$,
(for simplicity here we consider the case of the scalar order parameter,
i.~e. $n=1$) we have obtained the Dyson equation
\begin{equation} \label{eq:Dyson}
\frac{1}{2G({\bf k})}=r_0+ck^2
-\frac{\partial D(G)}{\partial G({\bf k})}
\end{equation}
where $D(G)$ denotes a quantity, the diagram expansion of which
involves all the so--called skeleton diagrams
(constructed of the fourth--order vertices \dshv with factors
$-V^{-1} u_{\bf k}$ related to the dashed lines) without outer lines.
Skeleton diagram is defined as a connected diagram containing no
parts like {\mbox \selfe.} In distinction to the usual (simple)
perturbation theory, the true correlation function $G({\bf k})$
is related to the coupled solid lines instead of the Gaussian
correlation function $G_0({\bf k})$. In such a way,
quantity $D(G)$ have to be considered as a
function of discrete variables $G({\bf k})$ corresponding to the set of
discrete wave vectors ${\bf k}$. Eq.~(\ref{eq:Dyson}) has
been previously obtained in Ref.~\cite{Ilmars}, where also
the simplest (ring) skeleton diagrams have been considered.
We have found a possibility to include all the skeleton
diagrams into consideration, which allows to find the set of
possible values for exact critical exponents.
At $u \to 0$ our equations, represented by converging sums and
integrals, define the true correlation function
$G({\bf k})$ with
an error smaller than $u^l$ at any positive $l$. This is quite enough to
find the exact critical exponents based on general scaling properties
of the solution in vicinity of the critical point. The structure of
our equations allows to prove these properties taking into account
all diagrams. The results also can be easily generalized
to the case with $O(n)$ symmetry. It is not possible in this
relatively short paper to give the mathematical derivations of our
equations and results. Hopefully, they will be available in nearest future.
Here we present the final result according to which possible exact
values of critical exponents $\gamma$ (susceptibility exponent) and
$\nu$ (correlation length exponent) for the $n$--component vector
model ($n$ = 1, 2, 3, etc.) in $d<4$ dimensions (i.~e., at $d=2, 3$) are
given by
\begin{equation} \label{eq:result}
\gamma = \frac{d+2j+4m}{d(1+m+j)-2j} \;; \hspace*{3ex}
\nu = \frac{2(1+m)+j}{d(1+m+j)-2j} \; ,
\end{equation}
where $m$ may have a natural value starting with 1 and $j$ is integer
equal or larger than $-m$. In general, different values of $j$ and $m$
can correspond to different (natural) $n$, i.~e., $j=j(n)$ and $m=m(n)$.
It is easy to verify that at $j=0$ and
$m=3$ Eq.~(\ref{eq:result}) reproduces the
known~\cite{Baxter} exact results in two dimensions.
The known exact exponents for the spherical model~\cite{Baxter}
($n=\infty$) are obtained at $j(n)/m(n) \to \infty$. Although the
derivations are true for $d<4$, Eq.~(\ref{eq:result}) provides
correct result $\nu=1/2$ and $\gamma=1$ also at $d=4$.
It is reasonable to consider $d$ as a continuous parameter. This
leads to the conclusion that $m=3$ and $j=0$ are the correct values
 for the case $n=1$ not only at $d=2$, but also at $d=3$. In the
latter case we have $\gamma=5/4$ and $\nu=2/3$. The nearest
values of $\gamma$ and $\nu$ provided by Eq.~(\ref{eq:result}),
e.~g., at $j=1$ and $m=3$ or at $j=1$ and $m=4$ are then the most
probable candidates for the case $n=2$.

\section{Comparison of results and discussion} \label{sec:compare}

It is commonly believed that all more or less correct Monte Carlo (MC)
simulations confirm the values of critical exponents obtained from the
perturbation expansions based on the renormalization group.
This is not true. We have found that
some kind of MC simulations at the critical point, namely,
 the MC simulations of fractal configurations of Ising model~\cite{IS}
and the MC simulations of the energy density~\cite{SM} for the $XY$
model in reality do
not confirm the results of the RG theory, but provide the values
of critical exponents which are very close to those we predicted.

\begin{figure}
\centerline{\psfig{figure=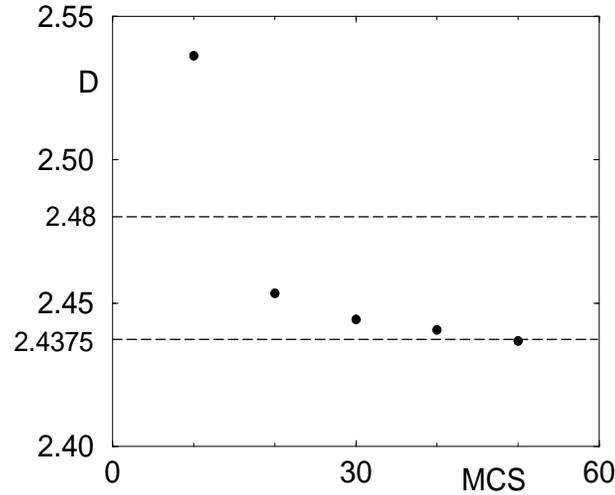,width=8cm,height=6.5cm,angle=-90}}
\caption{Fractal dimensionality $D$ of the three dimensional Ising
model at the critical point simulated by Monte Carlo method
(MCS means Monte Carlo steps). The upper and lower dashed lines
indicate the theoretical values expected from the known and from our
critical exponents, respectively.}
\label{fig1}
\end{figure}

 The MC simulations of Ref.~\cite{IS} allows to determine the fractal
dimensionality $D$ (the largest cluster in the relevant configuration
has the volume $L^D$ where $L$ denotes the linear size of the system)
which is related to the critical exponents by
$\gamma=\nu (2D-d)$ or, which is the same, $\eta=2-\gamma/\nu=d+2-2D$.
In our opinion, this method is better than other more convenient simulation
methods, since it provides the value of $\eta$ as a result of
direct simulation, i.~e., there are no fitting parameters. Besides,
the result is relatively insensitive to the precise value of the critical
coupling (temperature). In Fig.~\ref{fig1} we have shown the average
values of $D$ (the averaging is is made over the MC steps from 1 to 10
(except the initial point), from 11 to 20, and so on) calculated from the
MC data of Ref.~\cite{IS} by measuring deviation
from the line $D=2.48$ in Fig.~8 (of Ref.~\cite{IS}).
If properly treated,
these simulation data confirm the value of $\eta$ about $1/8$
(or $D=2.4375$) consistent with our prediction $\gamma=5/4$ and
$\nu=2/3$, as it is evident from Fig.~\ref{fig1}. The value
$D=2.46 \pm 0.01$ reported in Ref.~\cite{IS} seems to be determined
from the upper MC points (Fig.~8 in Ref.~\cite{IS}) only which are
closer to the known theoretical prediction $D=2.48$.

\begin{figure}
\centerline{\psfig{figure=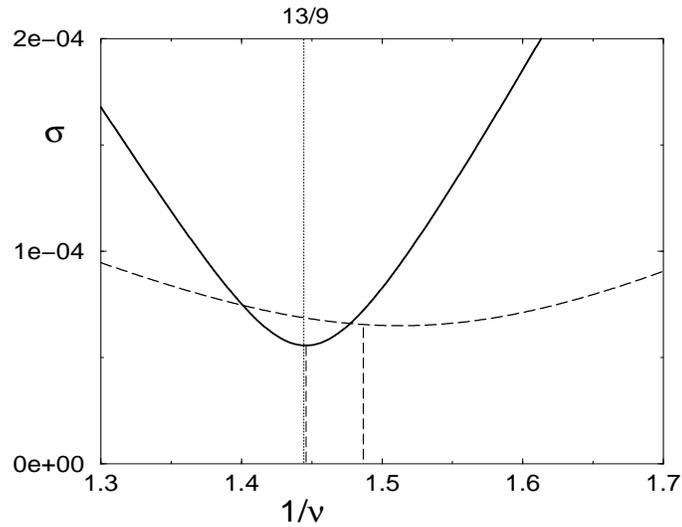,width=9cm,height=7cm,angle=-90}}
\caption{The standard deviation $\sigma$
vs the value of $1/\nu$ used in the least--square fit of
the finite--size scaling curve to the simulated results including
11 data points (solid curve) and 9 data points (dashed curve).
Minimum of the solid curve, shown by a vertical dashed line,
corresponds to the best fit $1/\nu=1.4457$ which is close to
our theoretical value $13/9$ indicated by a vertical dotted line.
Other vertical dashed line indicates the value $1.487$ proposed
by authors of Ref.~\cite{SM}. }
\label{fig2}
\end{figure}

  As regards the MC simulations
of the energy density $E$ of $XY$ model~\cite{SM} at the critical point,
the true picture can be reconstructed from the simulated values listed
in Tab.~I of Ref.~\cite{SM}. Since all the values of $E$ are of comparable
 accuracy, it is purposeful to use the least--square method to find the
optimum value of $1/\nu$ by fitting the MC data to the prediction of the
finite--size scaling theory
\begin{equation} \label{eq:anal}
E(L)=E_0+E_1 L^{{1 \over \nu}-d} \;,
\end{equation}
where $E(L)$ is the energy density at the critical temperature
$T_{\lambda}$ depending on the linear size of the system $L$.
The standard deviation of the simulated data points from the
analytical curve~(\ref{eq:anal}) can be easily calculated for
any given value of $1/\nu$ with the parameters $E_0$ and
$E_1$ corresponding to the least--square fit. The result is shown
in Fig.~\ref{fig2}.
The thick solid curve is calculated including all 11 data points
($L$=10, 15, 20, 25, 30, 35, 40, 45, 50, 60, 80), whereas
the dashed line -- including 9 data points (except $L$=10, 15)
used for the fitting in Ref.~\cite{SM}.
Minimum of the solid curve, shown by a vertical dashed line,
corresponds to the best fit $1/\nu=1.4457$ which comes very close to
our theoretical value $13/9$ ( provided by~(\ref{eq:result}) at
$j=1$ and $m=3$) indicated by a vertical dotted line. We have estimated
the statistical error of this MC result about $\pm 0.007$ by
comparing the best fits for several random data sets. Different data
sets have been generated from the original one by omitting some
data points with $10<L<80$. We have found it unreasonable
to omit the data points with two smaller sizes, as it has been
proposed in Ref.~\cite{SM}, since the result in this case becomes
very poorly defined, i.~e., the dashed curve in Fig.~\ref{fig2} has a
very broad minimum. Besides, there is no reason to omit the smallest
sizes, since the analytical curve~(\ref{eq:anal}) excellently fit
all the data points and the standard deviation for 11 data points
is even smaller than that for 9 data points (see Fig.~\ref{fig2}).
The possible systematical error
due to the inaccuracy in the critical temperature
$T_{\lambda}=2.2017 \pm 0.0005$ (the error bars are taken
from the source of this estimation~\cite{Janke}) used in the
simulations~\cite{SM} has been evaluated $\pm 0.017$ by comparing
the simulation results at $T_{\lambda}$ values  2.2012, 2.2017,
and 2.2022. In this case the values of the energy density at
a slightly shifted temperature have been calculated from the
specific heat data given in Tab.~I of Ref.~\cite{SM}.
In such a way, our final estimate from the original MC data
of Ref.~\cite{SM} is $1/\nu=1.446 \pm 0.025$ in a good agreement
with our theoretical value $13/9=1.444...$ and in a clear disagreement
with the usual (RG) prediction about $1.493$. One can only wonder
where the value $1.487$ proposed in Ref.~\cite{SM} comes from.
It does not correspond neither to the best fit for 11 data points
nor to that for 9 data points, as it is evident from
Fig.~\ref{fig2}. The values of $1/\nu$ and $\alpha/\nu$
cannot be determined independently from the discussed here
energy density data.
One of them have to be calculated from the scaling relation
$\alpha/\nu + d =2/\nu$. If authors of Ref.~\cite{SM} were able to
determine $1/\nu$ with $\pm 0.081$ accuracy, then they should be
able to find $\alpha/\nu$ with $\pm 0.162$ accuracy. In this aspect,
the estimate $\alpha/\nu = -0.0258 \pm 0.0075$ given by the
authors looks more than strange.

\begin{figure}
\centerline{\psfig{figure=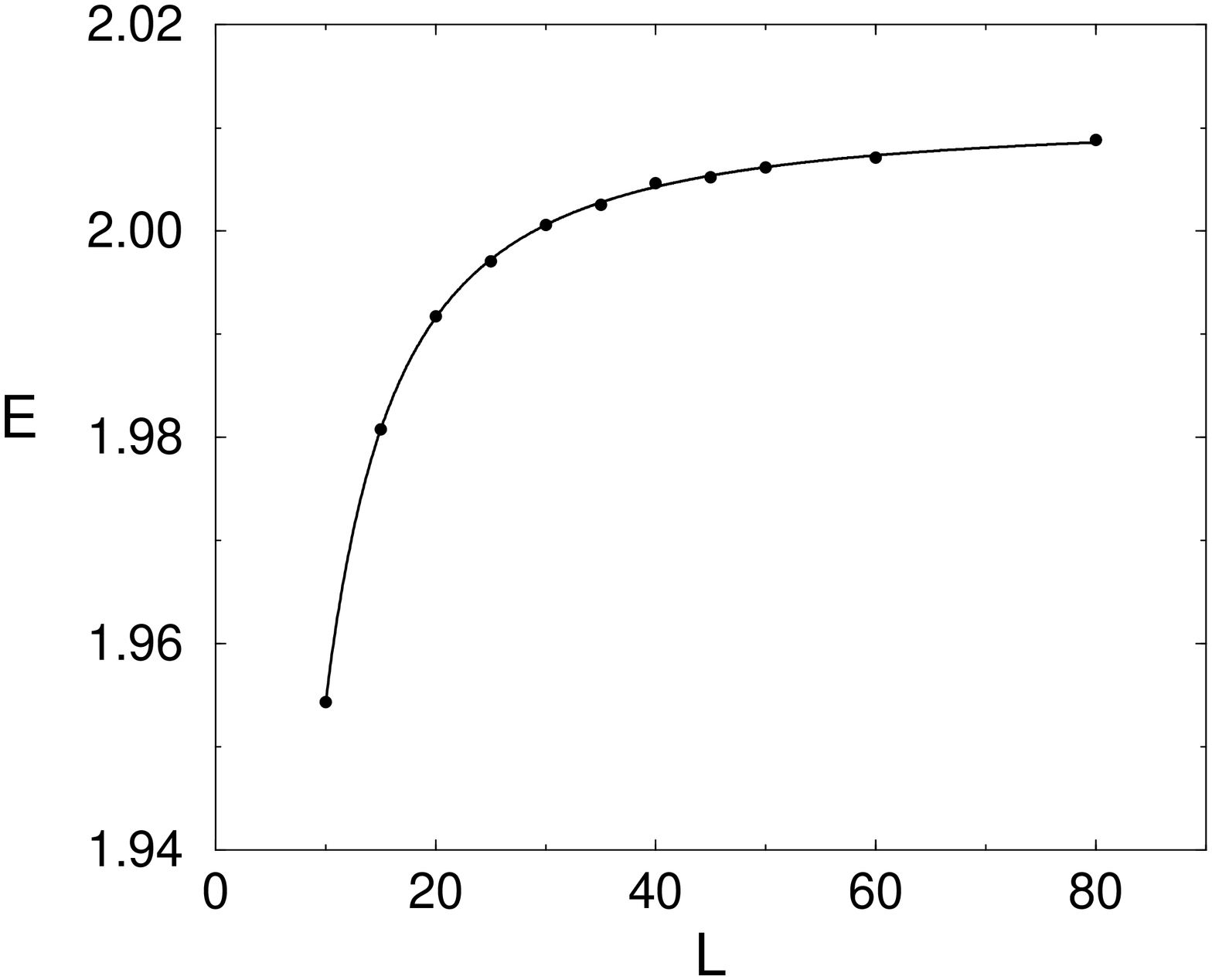,width=6.5cm,height=5.5cm,angle=-90}
\hspace*{1ex} \psfig{figure=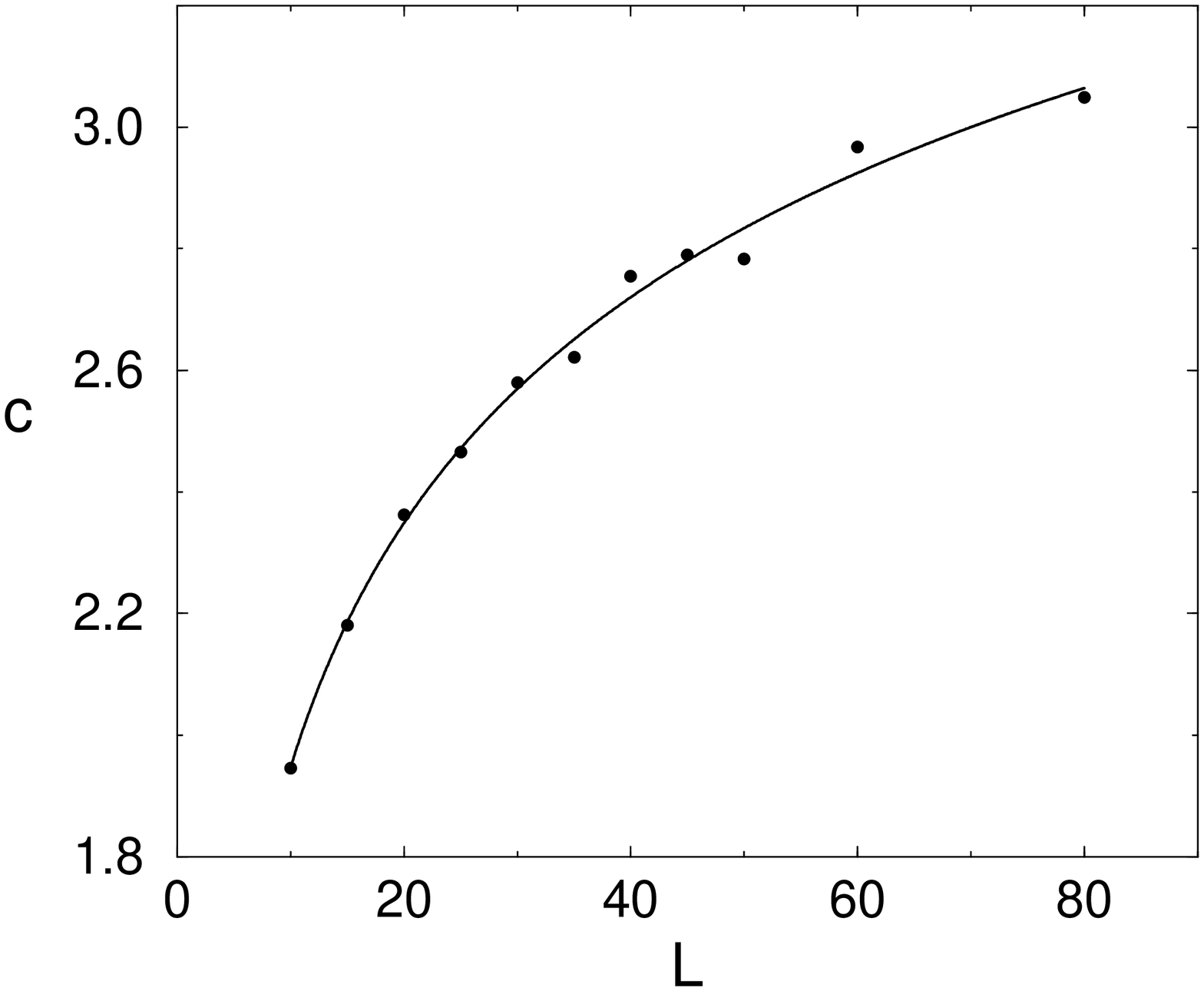,width=6.5cm,height=5.5cm,angle=-90} }
\caption{Our fits to the original MC data of Ref.~\cite{SM}
for the energy density (left) and for the specific heat (right)
depending on the linear size of the system $L$. }
\label{fig21}
\end{figure}

In Fig.~\ref{fig21} we have shown our fits to the MC data for the energy
density $E(L)=2.0108-2.0286 \, L^{-14/9}$
and for the specific heat $c(L)=7.360-6.990 \, L^{-1/9}$.
They do not look worse than those in Ref.~\cite{SM},
 but our fit for $c(L)$ seems to be better.

\begin{figure}
\centerline{\psfig{figure=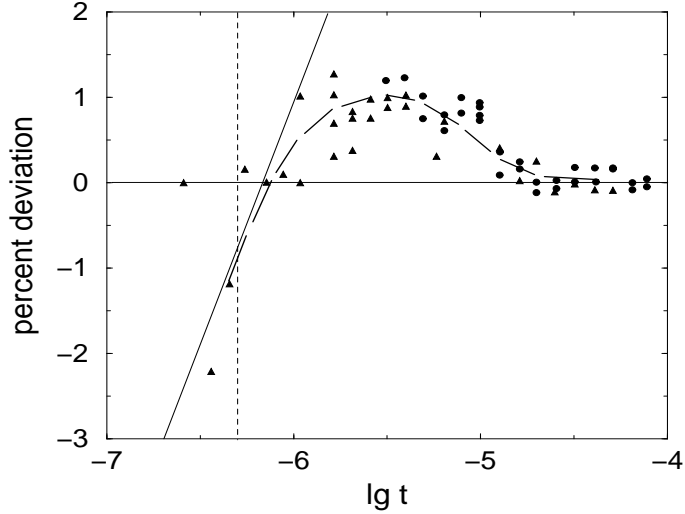,width=9cm,height=7cm,angle=-90}}
\caption{Percent deviation of the experimental $\rho_s /\rho$
data~\cite{GA} from the expected theoretical relation (Eq.~(\ref{eq:fit})
at $\delta=0$). The stright line shows the slope of this plot at
the value of $t$, equal to $5 \cdot 10^{-7}$, indicated by a vertical
dashed line. }
\label{fig3}
\end{figure}

  One believes that the value of critical exponent $\nu$
about $0.67$, predicted by the RG theory at $n=2$, is well
confirmed by very accurate measurements of the superfluid fraction
$\rho_s/\rho = y$ in $^4 He$. This is not true, since in reality
these experiments~\cite{GA}
provide a good evidence that the effective critical exponent
$\nu_{eff}(t)=\partial (\ln y)/ \partial (\ln t)$
remarkably increases when the reduced temperature
$t=(T_{\lambda}-T)/T_{\lambda}$ (where $T_{\lambda}$ is the
critical temperature) is decreased below $10^{-5}$.
According to Ref.~\cite{GA}, $\rho_s/\rho$ is given by
\begin{equation} \label{eq:fit}
\rho_s/\rho= y(t) =k_0(1+k_1 t) (1+D_{\rho} t^{\Delta}) t^{\zeta}
\times (1+\delta(t)) \; ,
\end{equation}
where $k_0$, $k_1$, $D_{\rho}$, and $\zeta$ are the fitting
parameters, $\Delta=0.5$ is supposed to be the correction--to scaling
exponent, and $\delta(t)$ is the measured relative deviation
from the expected theoretical expression obtained by setting
$\delta(t)=0$. The percent deviation discussed in Ref.~\cite{GA}
is 100 times $\delta(t)$. From Eq.~(\ref{eq:fit}) we obtain
\begin{equation} \label{eq:ffit}
\nu_{eff}(t)= \zeta + \frac{k_1 t}{1+k_1 t}
+ \frac{\Delta D_{\rho} t^{\Delta}} {1+D_{\rho} t^{\Delta}}
+ \frac{1}{1+ \delta(t)}
\times \frac{\partial \delta(t)}{\partial (\ln t)} \; .
\end{equation}
For the values of $t$ as small as $t<10^{-5}$ and for $\delta(t) \ll 1$
 Eq.~(\ref{eq:ffit}) with the fitting parameters $\zeta=0.6705$, $k_0=2.38$,
$k_1=-1.74$, and $D_{\rho}=0.396$ used in Ref.~\cite{GA} reduces to
\begin{equation} \label{eq:fi}
\nu_{eff}(t) \simeq \zeta + \partial \delta(t) / \partial (\ln t) \;.
\end{equation}
The second term in this equation is proportional to the slope of the
percent deviation plot $100 \, \delta(t)$ vs $\ln t$ or $\lg t$
(the decimal logarithm) in Figs.~2 and 3 of Ref.~\cite{GA}. We have read the
experimental data from Fig.~2 in Ref.~\cite{GA} within the region
$t<10^{-4}$ and have depicted them in Fig.~\ref{fig3}.
Almost all the data points
with a reasonable accuracy fit the smooth curve $\delta(t)$ vs $\lg t$
(dashed line) having a maximum at about $\lg t= -5.5$. It means that
$\partial^2  \delta(t) / \partial (\ln t)^2$ is negative within some
region around the maximum, i.~e., according to~(\ref{eq:fi})
the effective critical exponent $\nu_{eff}(t)$ increases if $t$ is decreased.
We have roughly estimated and have shown by  stright line the
slope of this curve at $t=t^* =5 \cdot 10^{-7}$ ($t^*$ value is indicated
in Fig.~\ref{fig3} by vertical dashed line). From this we obtain
$\partial \delta(t)/\partial(\ln t) \approx 0.025$.
This result depends on
the shift in the experimentally determined $T_{\lambda}$ value.
To obtain a more reliable estimate, we have performed the same manipulations
with the data depicted in Fig.~3 of Ref.~\cite{GA} corresponding to
$T_{\lambda}$ shifted by $\pm 20 nK$, and have obtained the values of
$\partial \delta(t)/\partial(\ln t)$ about $0.03$ and $0.015$, respectively.
Our final result $0.0233 \pm 0.0083$ for this derivative at $t=t^*$ has
been obtained by averaging over the three above discussed
estimates ($0.015$, $0.025$, and $0.03$) with the error bars large enough
to include all these values. According to this, from
Eq.~(\ref{eq:fi}) with $\zeta =0.6705$ we obtain
$\nu_{eff}(t^*)=0.694 \pm 0.009$ which, again, is in a good agreement with
the value $\nu=9/13 \simeq 0.6923$ provided by Eq.~(\ref{eq:result})
at $j=1$ and $m=3$ and in a disagreement with the RG predictions.

\section{Conclusions}

We have proposed a novel method (Sect.~\ref{sec:my}) which allows to predict
the exact values of critical exponents in the Ginzburg--Landau phase
transition model. Our proposal is accompanied by a
critical analysis of the conventional (perturbative) RG method.
In view of this analysis (Sect.~\ref{sec:RG} and \ref{sec:random})
and comparison with  MC simulation results and experiments
(Sect.~\ref{sec:compare}), our results should not be doubted from
the positions of the conventional (RG) theory. The best evidence of the
correctness of our treatment is the precise agreement with the known exact
solutions.

\end{document}